# Temperature-linear Resistivity in Twisted Double Bilayer Graphene

Yanbang Chu[1,2], Le Liu[1,2], Cheng Shen[1,2], Jinpeng Tian[1,2], Jian Tang[1,2], Yanchong Zhao[1,2], Jieying Liu[1,2], Yalong Yuan[1,2], Yiru Ji[1,2], Rong Yang[1,3], Kenji Watanabe[4], Takashi Taniguchi[5], Dongxia Shi[1,2,3], Fengcheng Wu[6], Wei Yang[1,2,3]*, Guangyu Zhang[1,2,3]*

[1] Beijing National Laboratory for Condensed Matter Physics, Institute of Physics, Chinese Academy of Sciences, Beijing 100190, China

[2] School of Physical Sciences, University of Chinese Academy of Sciences, Beijing, 100049, China

[3] Songshan Lake Materials Laboratory, Dongguan, Guangdong, 523808, China

[4] Research Center for Functional Materials, National Institute for Materials Science, 1-1 Namiki, Tsukuba 305-0044, Japan

[5] International Center for Materials Nanoarchitectonics, National Institute for Materials Science, 1-1 Namiki, Tsukuba 305-0044, Japan

[6] School of Physics and Technology, Wuhan University, Wuhan 430072, China

[7] Wuhan Institute of Quantum Technology, Wuhan 430206, China

*Corresponding authors: wei.yang@iphy.ac.cn; gyzhang@iphy.ac.cn

**Abstract**

We report an experimental study of carrier density ($n$), displacement field ($D$) and twist angle ($\theta$) dependence of temperature ($T$)-linear resistivity in twisted double bilayer graphene (TDBG). For a large twist angle ($\theta$>1.5°) where correlated insulating states are absent, we observe a $T$-linear resistivity (with the slope of the order ~10Ω/K) over a wide range of carrier density and its slope decreases with increasing of $n$, in agreement with acoustic phonon scattering model semi-quantitatively. The slope of $T$-linear resistivity is non-monotonically dependent on the displacement field with a single peak structure. For device with $\theta$~1.23° at which correlated states emerge, the slope of $T$-linear resistivity is found maximum (~100Ω/K) at the boundary of the halo structure where phase transition occurs, with signatures of continuous phase transition, Planckian dissipation, and the diverging effective mass; these observations are in line with quantum critical behaviors, which might be due to the symmetry-breaking instability at the critical points. Our results shed new light on correlated physics in TDBG and other twisted moiré systems.

**Main Text**

Resistivity ($\rho$) is a measure of how electrons are transported and scattered in solids, and it gives rich and fundamental information of the underlying system. It usually shows a power law dependence on temperature ($T$), $\rho \propto T^\alpha$, where α is an exponent that differs for different scattering mechanisms. Of particular interest is linear regime where $\alpha = 1$, since it could originate from conventional acoustic phonon couplings (Fig. 1a) [1], but also it might indicate unconventional processes that are rooted in quantum criticality [2-4] (Fig. 1b). The latter strange metal behaviors are observed in various correlated systems, for instance cuprates [5], iron-based compounds [6,7], heavy fermion systems [8,9], Kondo lattices [10], and frustrated lattices [11,12], where quantum critical points are found with the instabilities of order parameters. So far, most previous studies require multiple samples to obtain a single phase diagram, and improved in-situ sample tunability is demanded.

Moiré superlattices have emerged as a flat band system [13] for realizing a variety of interaction-driven quantum phases [14-19]. Large temperature linear (*T*-linear) resistivity has been observed in twisted bilayer graphene (TBG) [20,21], which bears a lot of similarity with that in optimally doped cuprates [22]. While the experiments in ref.20 support the electron-phonon scatterings [23,24], those in ref.21 raise the possibility of strange metal behavior with near Planckian dissipation in magic angle TBG, leaving the origin of the *T*-linear resistivity in TBG still elusive. In twisted double bilayer graphene (TDBG), observations of halo structures with field-tunable symmetry-breaking correlated states, as well as T-linear resistivity, have been reported [25-28]. The phase transitions from normal metallic states to correlated states occur at the boundaries of the halo structure in TDBG. Compared to TBG, TDBG is advantageous in tuning electron interactions in-situ by displacement field. The displacement field changes the flat bandwidth *W* and superlattice gap Δ in the band structure of TDBG [29], and thus acts as an extra parameter to control the relative strength of electron interactions to kinetic energy. The previous reports mainly focus on the nonlinear regime where *ρ* rapidly drops at low T, and the linear behaviors at high T are observed yet barely explored, demanding an in-depth and systematic exploration.

In this work, we systematically study *T*-linear resistivity in TDBG. We found *T*-linear resistivity in devices with twist angles from 1.23° to 1.91°. Firstly, we study the devices with large twist angles (*θ* >1.5°) where correlated insulating states are absent, and find that acoustic phonon scatterings can fully account for the *T*-linear resistivity. Meanwhile, we also demonstrate a displacement field tunable electron-phonon interaction. Secondly, in devices with *θ* ~1.23°, we reveal the features beyond the phonon model, and discuss the possibility of quantum criticality.

The TDBG samples are prepared by 'cut and stack' technique [30-33], and the devices are fabricated with a dual gate Hall bar geometry in bubble-free region of the samples. The dual gate geometry enables independent tuning of carrier density (*n*) and electric displacement field (*D*), via $n = (C_{BG}V_{BG} + C_{TG}V_{TG})/e$ and $D = (C_{BG}V_{BG} - C_{TG}V_{TG})/2\varepsilon_0$, where $C_{BG}$ ( $C_{TG}$ ) is the geometrical capacitance per area for bottom/top dielectric layer, $V_{BG}(V_{TG})$ is the bottom/top gate voltage, *e* is the elementary charge, and $\varepsilon_0$ is the vacuum permittivity. Fig. 2a shows ρ(*n, D*) at *T*=1.8K in 1.55° device, revealing the charge neutral point (CNP) and moiré gaps on electron branches (+n$_s$). The moiré filling factor $v = 4n/n_s$ corresponds to the number of carriers per moiré unit cell, where $n_s = 4/A \approx 8\theta^2/\sqrt{3}a^2$ is the carrier density at full filling and *a* is the graphene lattice constant.

We first focus on 1.55° device in Fig. 2 where correlated states are absent at *T*=1.8K. Fig. 2b shows the *ρ*(*T*) curves for fillings *v* from 0.5 to 2.5 at *D*=0 V/nm. *T*-linear resistivity with *ρ* ∝ *A$_1$T* is observed at *T*>*T\**, indicated by the yellow dashed lines in Fig. 2b. Here, *A$_1$* is the linear slope and *T\** is the characteristic temperature that separates linear regime at *T*>*T\** and nonlinear at *T*<*T\**. The $A_1$ is plotted against $n$ in Fig. 2c with values of about 10-30 Ω/K, and it decreases as $n$ increases; the onset temperature $T^*$ can be obtained in a quadratic plot $\rho-T^2$ (see Supplementary Material Fig. S3) [34], and it increases with $n$ as shown in Fig. 2d.

Our experimental data are captured quantitatively by electron-phonon scattering model in TDBG [35]. In this model, acoustic phonon scattering [36-39] is enhanced due to a reduced Fermi velocity in TDBG moiré bands, and it predicts a crossover from $T$ at high temperature to $T^4$ at low temperature. If electron-electron Umklapp scattering [40] dominates over phonon scattering at low temperature, the nonlinear term could become $T^2$, as shown by the black dashed lines in Fig. 2b. In the linear regime, the slope $A_1$ resulted from acoustic phonon scattering [35] is given by

$$A_1 = \frac{\pi D_A^2 k_B z_\infty}{2 g_s g_v e^2 \hbar \rho_m v_{ph}^2 v_F^2} \tag{1}$$

where $g_s (g_v)$ are spin(valley) degeneracies, $\rho_m$ is the mass density of graphene, $D_A$ is acoustic phonon deformation potential, $v_{ph}$ and $v_F$ are phonon velocity and Fermi velocity, $z_\infty$ is the integral concerning phonon occupation and other scattering details in TDBG. We plot the calculated $A_1$ as a function of $n$ for $\theta$=1.55° in Fig. 2c (orange dashed line), using $\rho_m$= 7.6×10$^{-7}$ kg/m$^2$, $v_{ph}$= 2.1×10$^4$ m/s, $D_A$ = 25 eV, $z_\infty$= 1.55 [20,35], and $v_F = \sqrt{(\hbar^2 n\pi/2)}/m^*$ with effective mass $m^* \approx$ 0.15$m_e$ calculated from band structure in continuum minimal model [13,41]. The calculated $A_1$ and the measured value in Fig. 2c are in good agreement. Moreover, the measured $T^*$ matches well with 1/4 of Bloch-Grüneisen temperature $T_{BG}$ in Fig. 2d, which further supports the phonon model [35]. Here $T_{BG}$ is defined as $2\hbar v_{ph} k_F / k_B$, where $k_F = \sqrt{\pi n/2}$ is the effective Fermi wave vector.

We also demonstrate that the electron-phonon scatterings are tunable by D field. Fig. 2e shows the $\rho(T)$ curves at $\nu = 2$ for different field $D$ from 0 to -0.7V/nm in the 1.55° device, with the linear behaviors well preserved. Fig. 2f presents the curves of extracted $A_1$ as a function of $D$ at different $\nu$, which shows a non-monotonical dependence with peaks at finite $D$. Since $A_1 \propto m^{*2}$ at a fixed $\nu$, the $D$-dependent $A_1$ suggests a field tunable moiré band dispersion in TDBG [42]. As indicated by the arrow in Fig. 2f, the peak position in $D$ shifts with $\nu$, demonstrating the electron filling effect on band structure [26,43].

Next, we turn to the 1.23° device with correlated insulating states and concomitant halo structure in moiré conduction band. $T$-linear resistivity with $\rho \sim A_1 T$ is observed at $\nu$=2 for $T > T^*$ in Fig. 3a-b. Sublinear behaviors at higher temperatures are also observed, which might be due to thermal excitations of remote dispersive bands [20]. The extracted $A_1$ is plotted against the $D$ field as red dots in Fig. 3c, and it is strongly field tunable with a large magnitude of ~100Ω/K. The non-monotonical dependent $A_1(D)$ shows two peaks with an "M" shaped structure in Fig. 3c, and the peaks locate at the critical point of metal-insulator transition. The "M" shaped two peaks of $A_1$ are also observed at $\nu$=1.8 (Fig. 4a) and other fillings (see Supplementary Material Fig. S6) [34]. At $\nu$=1.8, the $A_1$ peaks are found concurrent with a sharp change of Hall coefficient, indicating phase transitions between metallic phases with broken symmetry and those without [28]. Moreover, these $A_1$ peaks are found following the boundary of the halo structure in Fig. 4d, implying the correlation between enhancement of $A_1$ and phase transitions. From the phonon model, the increase of $A_1$ at the halo boundary suggests the enhancement of effective mass.

To better reveal the effective mass, we analyze the temperature dependence of resistivity at $T<T^*$. For the metals outside the halo structure, we observe signatures of quadratic resistivity $\rho \sim A_2 T^2$ at $T < T^*$, as shown in Fig. 3a. The $D$ dependent $T^*$ and $A_2$ are plotted in Fig. 3d-e. We find a decreasing $T^*$ and an increasing $A_2$ when approaching the critical displacement field. The quadratic power law at low temperature indicates Landau Fermi liquid behavior, in which $A_2 \propto k_F/E_F^2 \propto m^{*\alpha}/n^\beta$ with α and β being the exponents that depend on the details of band structure [44]. It is worth noting that the $A_2(D)$ also qualitatively agrees with the $A_1(D)$ in the phonon scatterings model, since $A_1 \propto 1/v_F^2 \propto m^{*2}/n$. The increased m* at the halo boundary is also revealed in the carrier density dependent $A_2$(n) at a fixed $D$ (Fig. 4c). Such quadratic dependence of resistivity is also well reproduced from another TDBG device with $\theta$ =1.28° in Fig.5a and 5b, and the corresponding field dependent $A_2$ also tends to diverge at the halo boundary in Fig. 5c, similar to those observations in Fig. 4. For reference, direct measurement of $m^*$ by Lifshitz-Kosevich method is also carried out for the 1.28° device in Fig. 5d-f.

One important question is why the m* tends to diverge at the boundaries of halo structure. Though the change of m* could explain the change of $A_1$ and $A_2$ based on electron-phonon scattering model and Fermi liquid scenario, the observation of diverging $m^*$ is unusual by itself, and it may call for the possibility of quantum criticality. Given the correlation between diverging $m^*$ and symmetry breaking phase transition at the critical points of halo structure, the diverging $m^*$ suggests the possible existence of quantum critical behaviors [45,46], which we discuss in the following. Firstly, the phase transitions at the critical points of the halo structure are continuous in the temperature range we studied, indicated by the vanishing thermal activation gap (Fig. 3f). Secondly, the decreasing $T^*$ and the increasing $A_2$ when approaching the critical displacement field outside the halo structure in Fig. 3d and Fig. 4b are also signatures of the quantum critical behaviors [45]. Lastly, we find the large T-linear resistivity falls into the Planckian dissipation regime where the scattering rate is proportional to temperature, $\hbar/\tau = Ck_B T$ and the coefficient $C$ approaches O(1). In graphene moiré system [21], the number $C$ can be extracted from T-linear resistivity slope $A_1$ by $C = \hbar e^2 n A_1 / k_B m^*$. In our cases, we obtain the $C \sim 1.8$ for the $A_1$ peaks at v=2 in Fig. 3c, by taking $m^*$=0.3$m_e$, n=1.75×10$^{12}$/cm$^2$, $A_1$=146Ω/K. All these facts suggest the possible existence of quantum criticality at the boundaries of halo structure.

At continuous phase transition critical points, quantum fluctuations associated with symmetry-breaking order parameters can be significant. In TDBG, the boundary of the halo structure separate states with different symmetry, which is likely to generate quantum fluctuations. More specifically, we suspect it is the spin fluctuations [47] that contribute to the T-linear resistivity. While spin-up electrons and spin-down electrons are equally filled outside the halo, one is more favored than the other inside the halo structure [26]. However, more experimental and theoretical investigations are required for a better understanding.

In conclusion, we systematically investigate the carrier density, displacement field and twist angle dependences of T-linear resistivity in TDBG. We demonstrate a dominant role played by acoustic phonons when correlated states are absent at $\theta$>1.5°. The T-linear resistivity has a non-monotonic displacement field dependence with a single $A_1$ peak, revealing the field-tunable electron phonon interaction in TDBG. Moreover, we observe a "M"-shaped two peak structure in the presence of correlated states at $\theta \sim 1.23°$. These peaks are found located at the halo boundary, separating states with different symmetries. We propose the possible existence of quantum criticality, supported by the evidences of continuous phase transition, vanishing $T^*$, Planckian dissipation, and the diverging effective mass at the critical points. Our observations establish a close link between high temperature $T$-linear resistivity to low temperature ground states, and hopefully inspire more works about the nature of quantum criticality and ground states instability in TDBG [48]. Similar phenomena may also be expected in other field-tunable systems such as twisted monolayer-bilayer graphene [49-51] and ABC trilayer graphene/hBN moiré system [52].

**Acknowledgment**: We thank Rui Zhou, S. Das Sarma, Xi Dai, Hongming Weng, Jianpeng Liu, Lede Xian, Xiao Li for helpful discussions. This work was supported by the National Key Research and Development Program (Grant No. 2020YFA0309600), the NSFC (Grants Nos. 61888102, 11834017, and 12074413), the Strategic Priority Research Program of CAS (grant Nos.XDB30000000 and XDB33000000), the Key-Area Research and Development Program of Guangdong Province (Grant No.2020B0101340001), and the Research Program of Beijing Academy of Quantum Information Sciences under Grant No. Y18G11. F. Wu is supported by National Key R&D Program of China 2021YFA1401300 and start-up funding of Wuhan University. Growth of hexagonal boron nitride crystals was supported by the Elemental Strategy Initiative conducted by the MEXT, Japan, Grant Number JPMXP0112101001, JSPS KAKENHI Grant Number JP20H00354 and A3 Foresight by JSPS.

**Author contributions**: W.Y. and G.Z. supervised the project; Y.C., C.S., W.Y., G.Z. conceived the experiments; Y.C., L.L., C.S. fabricated the devices and performed the transport measurements; Y.C., L.L., J.P.T. provided the samples; L.L. provided continuum model calculations; K.W. and T.T. provided hexagonal boron nitride crystals; Y.C., W.Y., G.Z. analyzed the data; Y.C. and W.Y. wrote the paper. All authors discussed and commented on this work.

# Figures and Figure captions

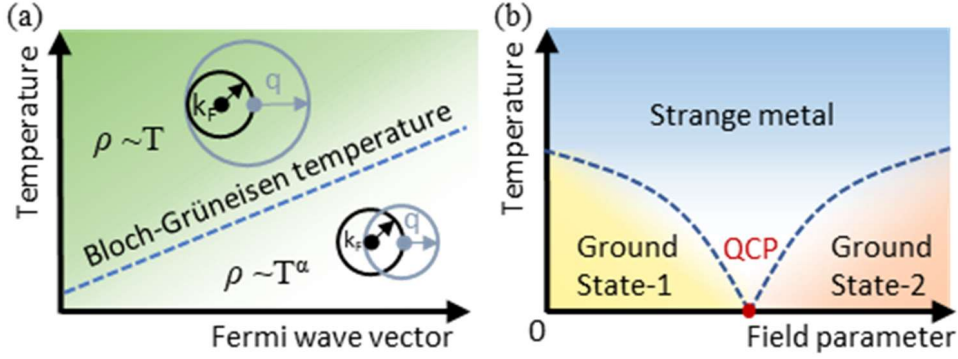

**Figure 1. Schematics of *T*-linear resistivity originated from (a) electron-phonon scattering and (b) quantum critical point (QCP).** The Bloch-Grüneisen temperature ($T_{BG}$) is defined when Fermi momentum $k_F$ is half of the maximum phonon wave vector $q$, and $\alpha>1$ at $T<T_{BG}$.

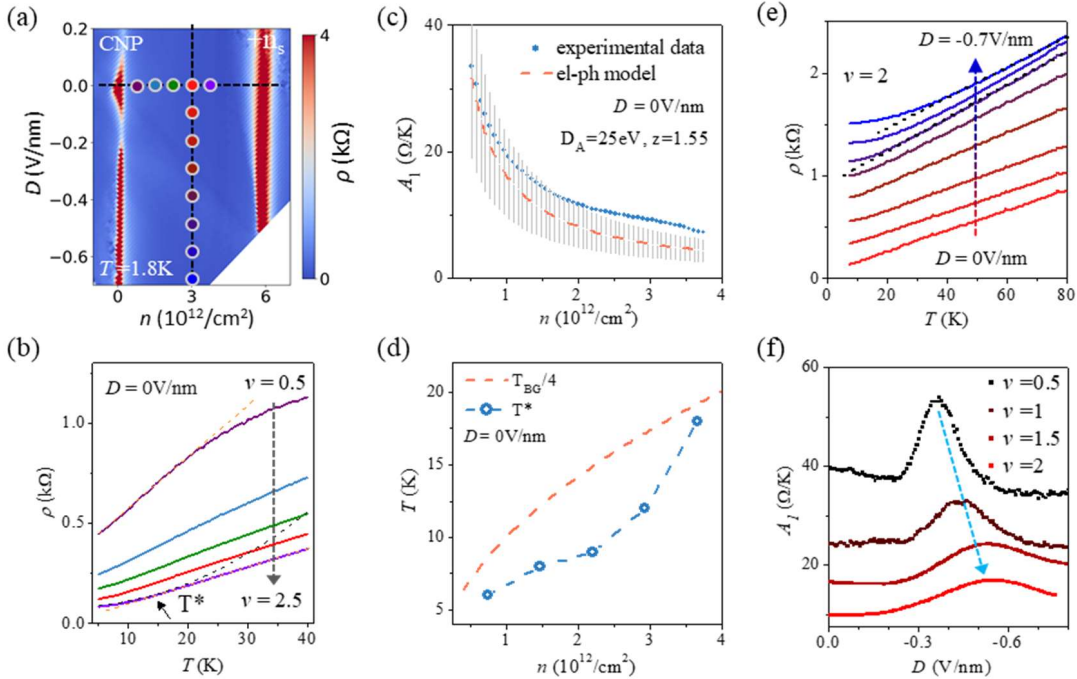

**Figure 2. *T*-linear resistivity and field-tunable electron-phonon interactions in 1.55° TDBG.** **(a)** A color mapping of $\rho(n,D)$ at $T=1.8$K. **(b)** $\rho(T)$ at $D=0$V/nm for $v$ from 0.5 to 2.5, with a step of 0.5. The black arrow marks the $T^*$. **(c)** Experimental $A_1(n)$ at $D=0$V/nm (blue solid line), and phonon model prediction (orange dash line) with $m^*=0.15m_e$ calculated from continuum minimal model. We set the error bar of the simulation to 40%, mainly due to the uncertainty of acoustic phonon deformation potential and effective mass. **(d)** 1/4 of theoretical Bloch-Grüneisen temperature ($T_{BG}/4$) and experimental *T*-linear onset temperature ($T^*$) for 1.55° device at $D=0$V/nm. **(e)** $\rho(T)$ at $v=2$ for $D$ from 0 to -0.7V/nm, with a step of 0.1V/nm and offset 200Ω between curves. The positions of these $\rho$-$T$ curves in **(b)** and **(e)** are marked as colored dots in **(a)**. **(f)** $A_1(D)$ at $v=0.5, 1, 1.5$ and 2, with an offset of 5Ω/K for clarity. The light blue dash arrow illustrates the shift of $A_1$ peak with decreasing $v$.

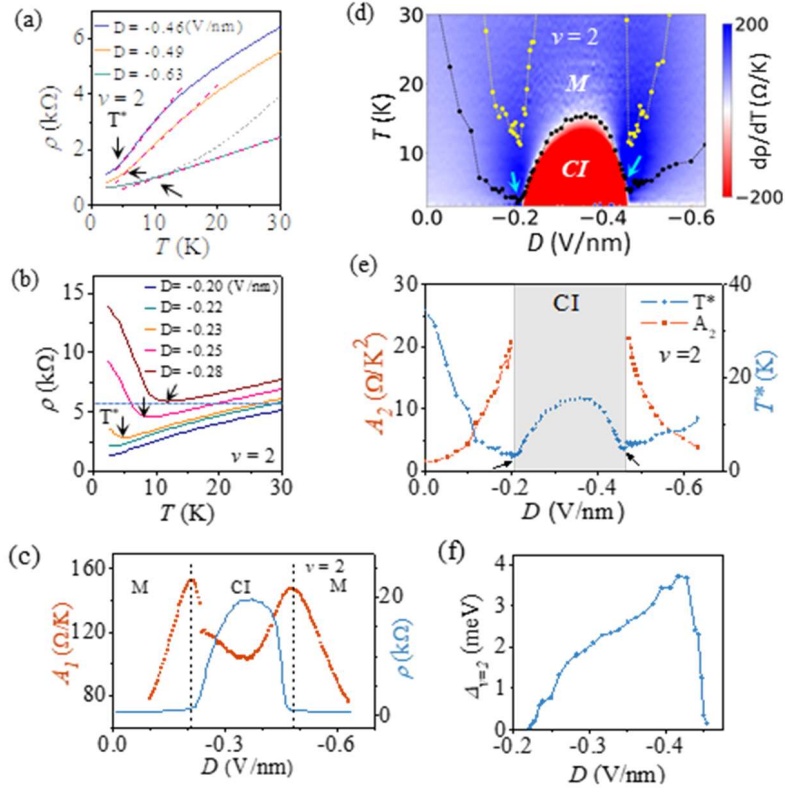

**Figure 3. *T*-linear resistivity and critical behaviors at the boundaries of halo structures in 1.23° TDBG.** **(a)** Plots of $\rho(T)$ outside the halo at *v*=2. The dash line is $T^2$ dependence fitted by experimental curve at *T*<*T**. The pink lines show the T-linear resistivity at *T*>*T**. **(b)** Plots of $\rho(T)$ near critical points at *v*=2. **(c)** Plots of $A_1(D)$ and $\rho(D)$ at *v*=2, where $\rho(D)$ is measured at *T*=1.8K. Here "M" denotes "conventional metal". **(d)** A mapping of the numerical d$\rho$/dT at *v*=2 **(d)**. The *T** are marked by black dots. The upper temperature boundaries of *T*-linear resistivity are marked by yellow dots. The cyan arrows indicate the critical points. **(e)** *D*-dependent $A_2$ and *T** at *v*=2. **(f)** Thermal excitation gap at *v*=2 as a function of displacement field.

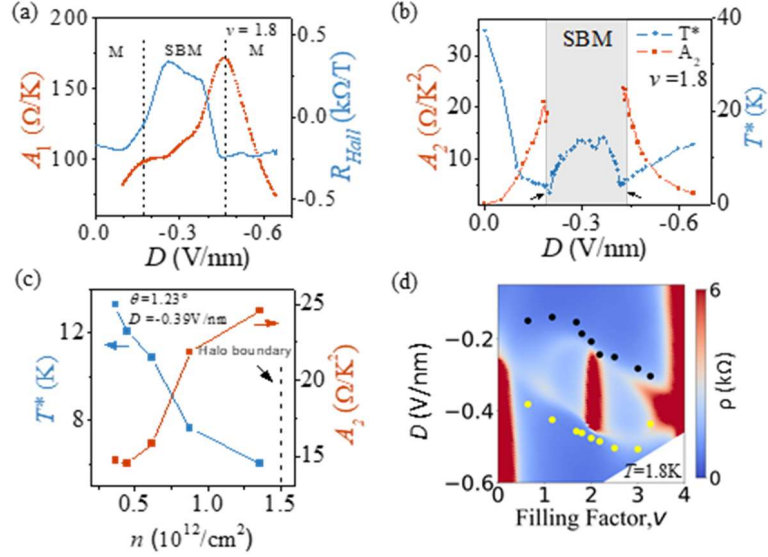

**Figure 4. *T*-linear resistivity near halo boundaries in 1.23° TDBG. (a)** $A_1$ and Hall coefficient as a function of displacement field. $R_{Hall}$ is measured at *B*=0.1T. The vertical dash lines mark the positions of $A_1$ peaks, which locate near the abrupt change of $R_{Hall}$. **(b)** *D*-dependent $A_2$ and $T^*$. The black arrows indicate the critical points, and the grey areas are the regime of correlated ground states inside the halo. **(c)** Quadratic temperature dependence of resistivity at constant displacement field in 1.23° TDBG. $T^*$ ($A_2$) decreases (increases) as approaching the halo boundary, consistent with the observation at constant carrier density measurement in Fig.3. **(d)** The positions of $A_1$ peaks for different *v* in the $\rho(v, D)$ mapping, marked as colored dots.

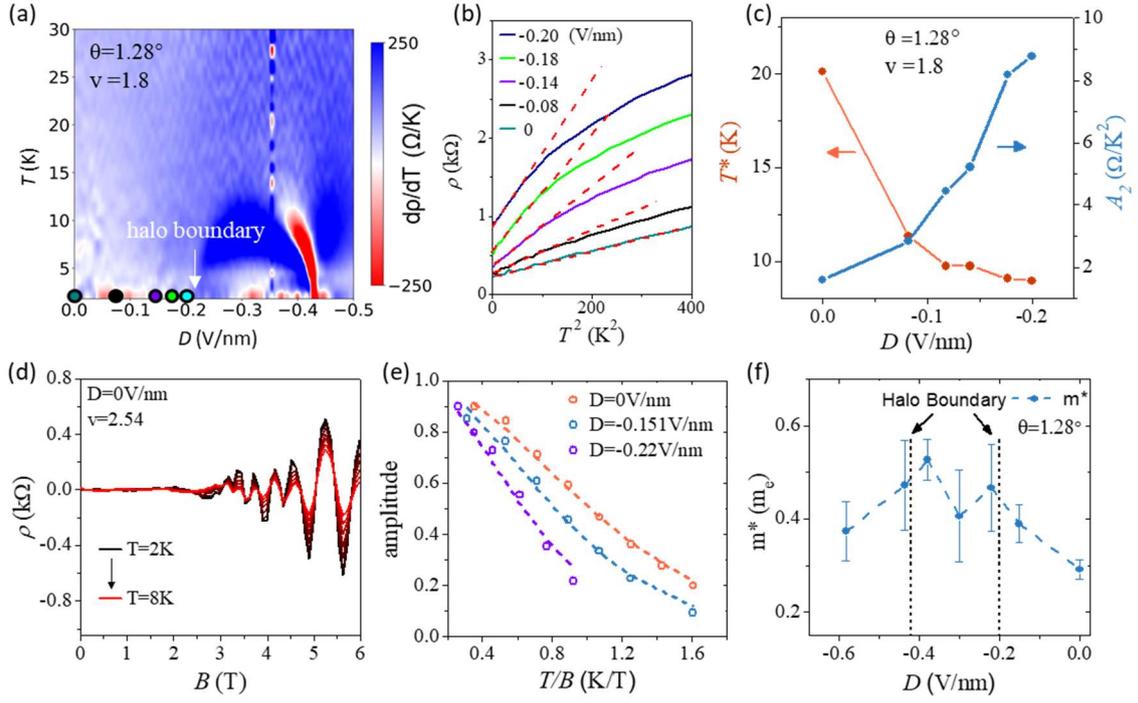

**Figure 5. Quadratic-in-*T* resistivity in 1.28° TDBG. (a)** a color mapping of d$\rho$/d*T* as a function of displacement field and temperature. **(b)** Resistivity as a function of $T^2$ at *v*=1.8. The positions of the displacement fields are marked by color dots in (a). The red dash lines show the increasing $A_2$ as approaching the boundary of the halo. **(c)** the field dependence of $T^*$ and $A_2$ extracted from $\rho$-$T^2$ plots in (b). **(d)** Temperature-dependent sdH oscillation amplitude. **(e)** Normalized oscillation amplitude as a function of *T*/*B* at representative displacement fields. **(f)** Extracted effective mass at different displacement fields.

# Supplementary Information

# I. Methods

**Sample fabrications.** Bilayer graphene flakes are exfoliated from highly oriented pyrolytic graphite, with the layer number determined by optical contrast and further confirmed by atomic force microscope. Thermal annealing is performed to ensure the stacking order of bilayer graphene to be Bernal stacking and remove contaminations on exfoliated graphene and hBN surfaces. Samples are assembled by "cut-and-stack" method technique using Poly Bisphenol A carbonate film coating on a polydimethyl siloxane stamp placed on a glass slide. The twist angle is controlled by a rotator stage with accuracy 0.02°. After the pick-up procedure, the obtained heterostructure is transferred onto silicon slice with 300nm silicon dioxide. Heterostructures are fabricated into Hall bar devices through electron beam lithography and reactive ion etching with $CHF_3/O_2$ as reactive gas, and contacted by Cr/Au through electron beam evaporation system. Heavily doped silicon or few layer graphite are used as bottom gate and Ti/Au metal film as top gate, respectively.

**Electrical measurements.** Devices are measured in a helium-4 vapor flow cryostat (1.8K) with a standard lock-in technique. Typical a.c. current is applied by Stanford SR830 with a low frequency of 30.9Hz and a current of 10nA. Temperature is controlled by Cryo Con 32B PID temperature controller with long enough delay time to ensure thermal equilibrium in the device.

**Linear fit procedures.** In the correlated regime, the linear range is limited and the linear fit is done at a temperature above the phase transitions. Firstly, we identify the critical temperature T* that separates correlated states and normal metal inside the halo structure. As shown in Fig. 3d and 4c, T* is also a good indicator as the onset temperature for the T-linear resistivity, i.e. ρ(T) shows a good linearity at T>T*. Then, as the field is tuned away from the center of the halo regime, T* gradually gets smaller and smaller, and it eventually approaches our base temperature at the boundary of halo structure. Once across the boundary, T* separate the regimes between the T-linear and quadratic power law, and it gradually grows when D is tuned further away from the halo regime. Finally, we end up in a situation where the linear temperature ranges near critical point and those far away are not overlapping.

**Electron-phonon scattering model.** From continuum model calculations, we get conduction moiré bandwidth (BW) and effective mass at different twist angles. Then, by $v_F = \sqrt{(\hbar^2 n\pi/2)}/m^*$, and eventually obtain the Fermi velocity. The electron phonon scattering model is programmed on Python, using $\rho_m$=7.6×10$^{-7}$kg/m$^2$, $v_{ph}$=2.1×10$^4$m/s, $D_A$=25eV, $z_\infty$=1.55.

**Others.** For the field dependent $A_1$ at a filling close to v=2, the peaks can be identified directly from the "M" shaped plots. For the field dependent $A_1$ at a filling away from v=2, one peak is prominent while the other is very faint, giving a shoulder like structure. We identify the faint one by subtracting a linear background of the shoulder feature. The linear background is defined by linking the two side-turning-points of the shoulder.

## II. Extended data

**Note A: Supplementary information for electron-phonon interaction in TDBG.**

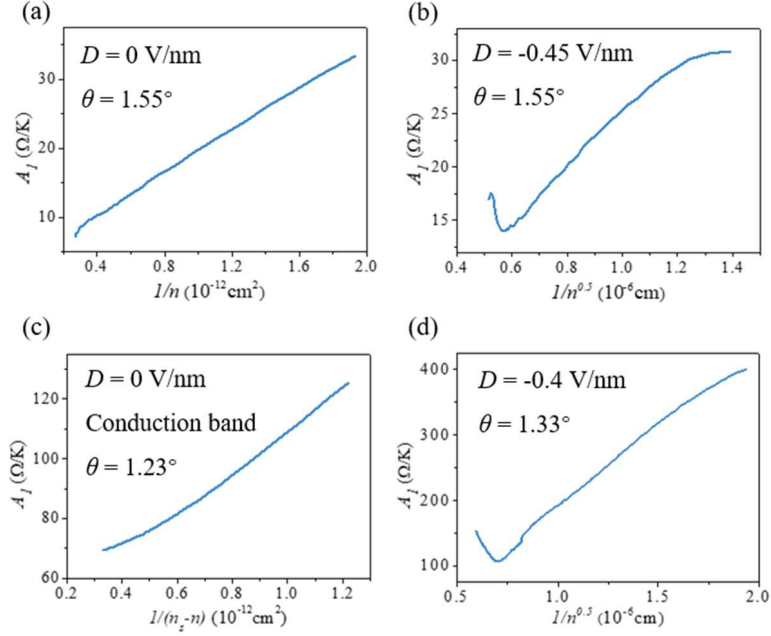

**Figure S1. Carrier density scaling of $A_1$.** The $A_1$ scales linearly with $1/\sqrt{n}$ in **(b)**, while scales with $1/n$ at $D=0$V/nm in 1.55° device in **(a)**. The different scaling of $1/n$ or $1/\sqrt{n}$ may root in field tunable band structure in TDBG. For ~1.3° device, **(c)** at $D=0$V/nm, the $A_1$ is linear in $1/n$ as well in both CB and VB. Here, $n$ is relative value to full filling of moiré conduction ($+n_s$) and valence ($-n_s$) band, respectively. At $D=-0.39$V/nm. **(d)** $A_1$ obtained from 1.33° device at the similar displacement field shows $1/\sqrt{n}$ scaling of $A_1$. The scaling other than $1/n$ rule suggests non-parabolic band at finite displacement field of TDBG.

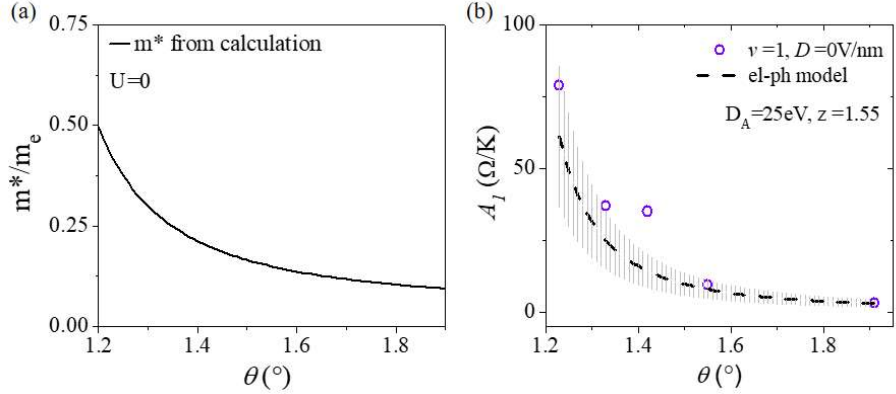

**Figure S2. Twist angle dependence of effective mass and electron-phonon model simulation.** **(a)** Effective mass extracted from band structure calculation at U=0meV by continuum minimal model. **(b)** Comparison between experimental $A_1$ and el-ph simulations at different twist angles. The dark blue dash line shows the el-ph simulations with calculated effective mass from (a). Here, $z=1.55$, $D_A=25$eV. The el-ph simulation error bar is 40%.

## Note B: Quadratic power law and Fermi liquid behaviors

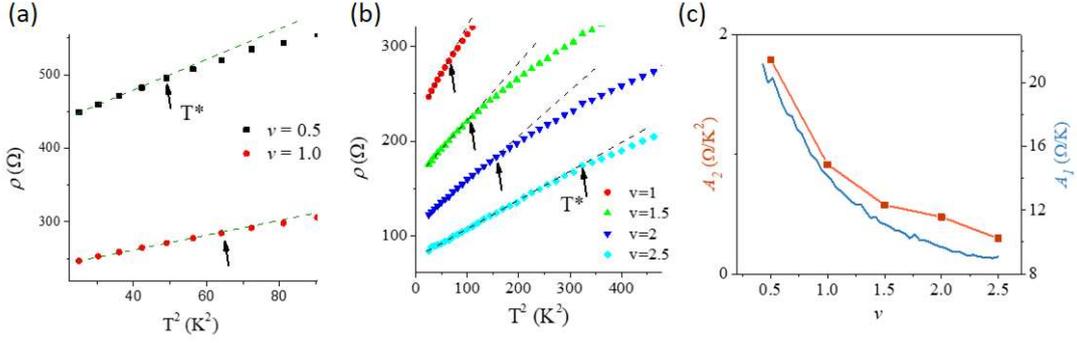

**Figure S3. Quadratic-in-T resistivity at low temperature in 1.55° TDBG.(a)** Resistance versus the square of temperature at $v$=0.5 and 1. The black arrows point to the $T^*$. **(b)** Resistance versus the square of temperature at $v$=1, 1.5, 2 and 2.5. The black arrows point to the $T^*$. **(c)** The comparison between $A_2$ and $A_1$ for $D$=0V/nm, 1.55° TDBG.

## Note C: Enhanced T-linear resistivity near the halo boundary.

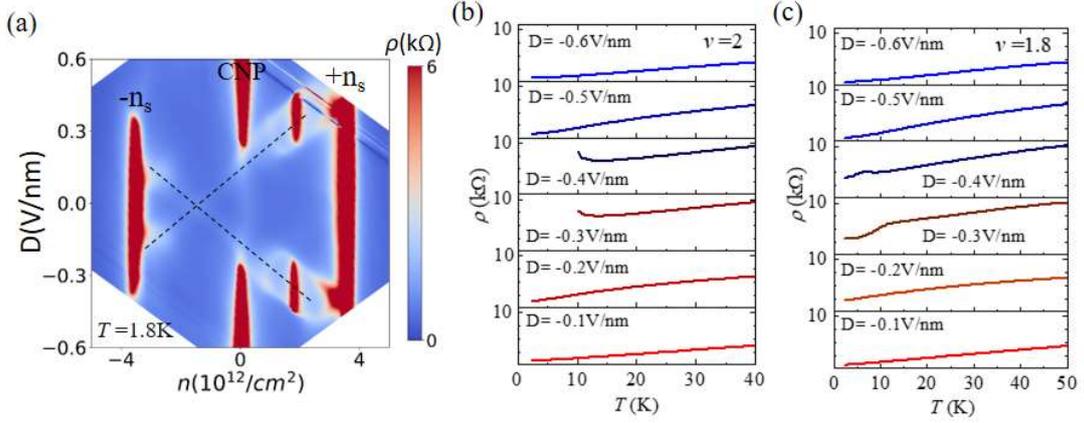

**Figure S4. T-linear resistivity in 1.23° TDBG.(a)** Mapping of $\rho(n,D)$ at $T$=1.8K. **(b-c)** Line cuts of $\rho(T)$ at $v$=2 **(b)** and $v$=1.8 **(c)** with $D$ changes from -0.1 to -0.6V/nm.

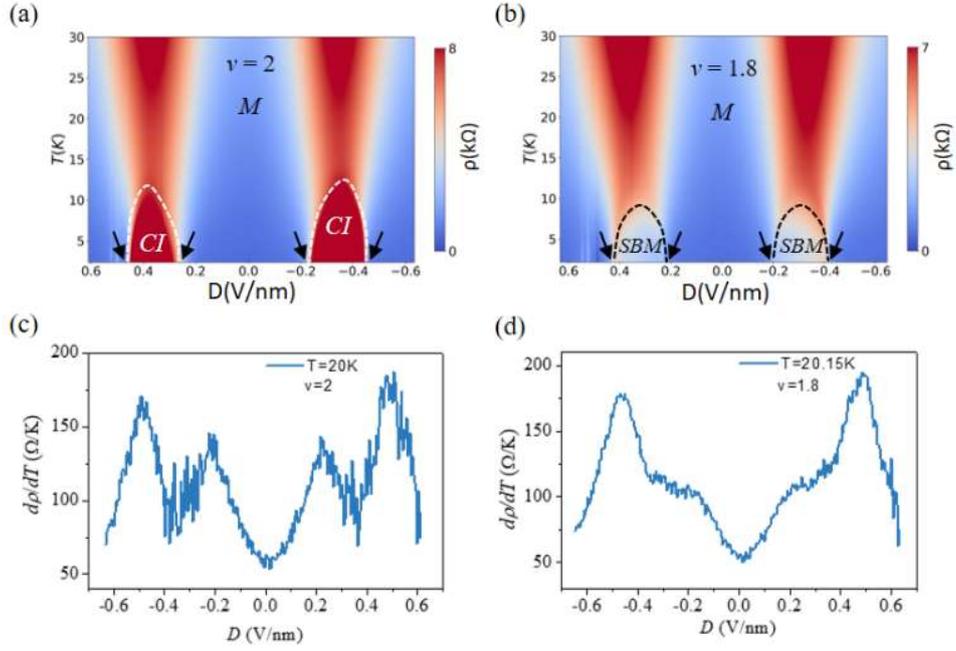

**Figure S5. Local numerical differential dρ/dT in 1.23° TDBG.(a-b)** Mappings of ρ(D,T) at *v*=2 **(a)** and *v*=1.8 **(b)**. 'CS','M' and 'SBS' refer to correlated insulating states, conventional metal, symmetry broken states, respectively. **(c-d)** Numerical dρ/dT as a function of *D* at *T*=20K for *v*=2 **(c)** and *v*=1.8 **(d)**. It's clear that 'M' shape feature and shoulder-like feature preserves. The numerical differential can rule out effect of averaging by linear fit, confirming the experimental observations in the main text.

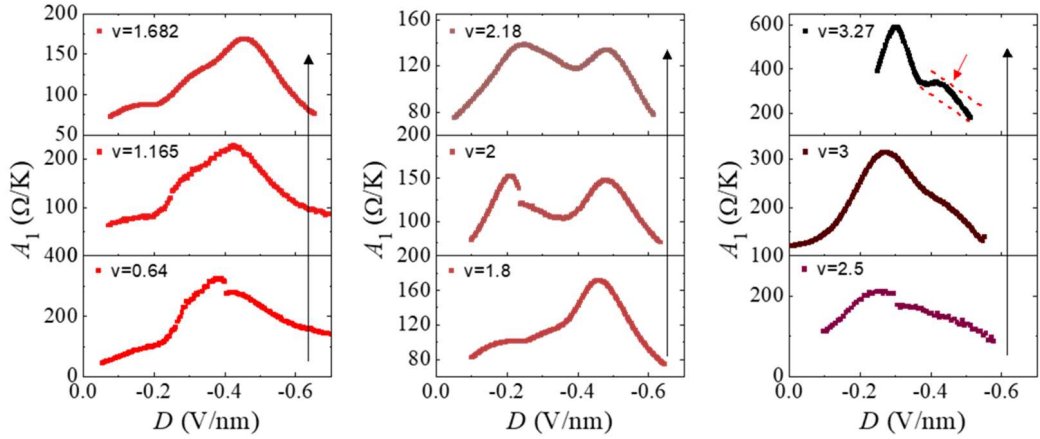

**Figure S6. $A_1$ as a function of displacement field at various *v* in 1.23° TDBG.** The orange dashed lines at the subfigure of *v*=3.27 are used to identify the faint peak (indicated by the orange arrow) from the shoulder structure.

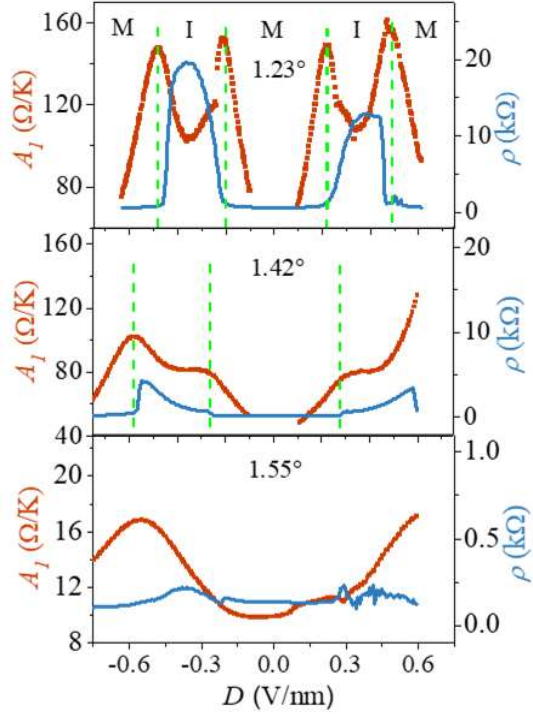

**Figure S7. Twist angle dependence of $A_1$ as a function of $D$.** $A_1$ and resistivity measured at base temperature as a function of displacement field at $\theta$=1.23° (top panel), 1.42° (middle panel), 1.55° (bottom panel) at $\nu$=2. Link between the 'M' feature and correlated insulating states can be confirmed.

**Note D: Temperature-dependent data in other devices.**

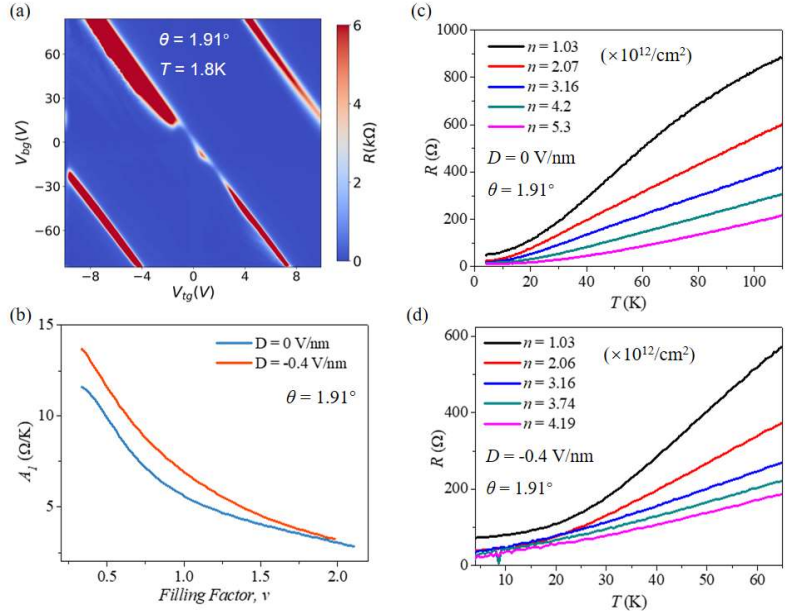

**Figure S8. T-linear resistance behavior in 1.91° TDBG.** **(a)** Resistance as a function of $V_{tg}$ and $V_{bg}$. **(b)** $A_1$ as a function of $\nu$ at $D$=0V/nm (dark blue) and $D$=-0.4V/nm (orange). **(c-d)** Resistance as a function of $T$ at $D$=0V/nm **(c)** and $D$=-0.4V/nm **(d)**.

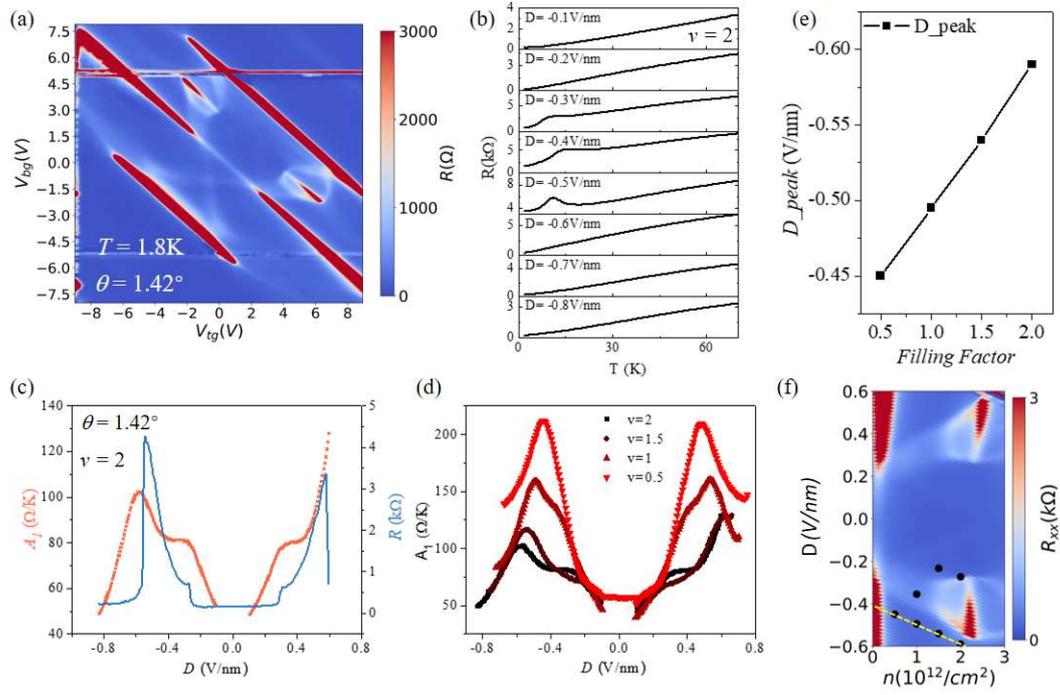

**Figure S9. T-linear resistance behavior in 1.42° TDBG.** **(a)** Resistance as a function of $V_{tg}$ and $V_{bg}$. **(b)** Resistance as a function of $T$ at $v=2$ with $D$ step of 0.1V/nm. **(c)**. $A_1$ and resistance as a function of $D$ at $v=2$. **(d)** $A_1$ as a function of $D$ at various $v$. **(e)** The displacement field of the $A_1$ peak as a function of $v$. **(f)** Zoom in of the resistance as a function of $v$ and $D$ in 1.42° device. Black dots mark the position of the $A_1$ peak, yellow dash line is the linear guide line for the eye.

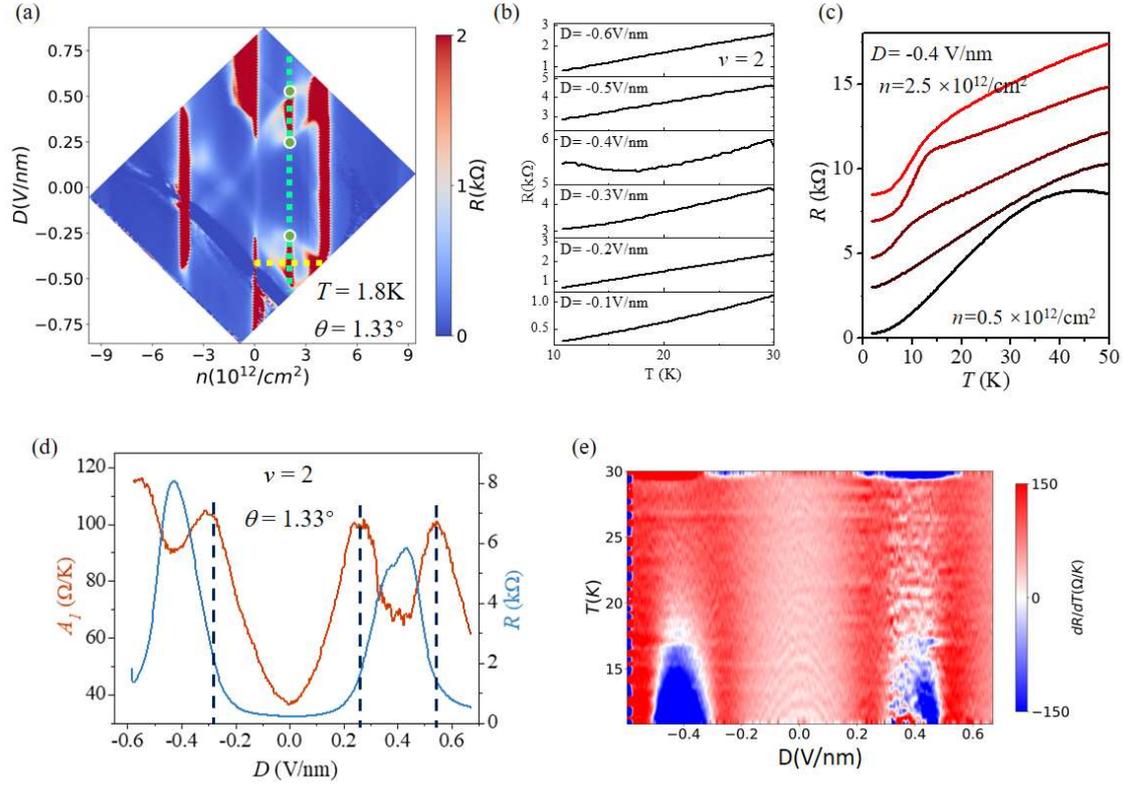

**Figure S10. T-linear resistance behavior in 1.33° TDBG. (a)** Resistance as a function of carrier density and displacement field in 1.33° device. **(b-c)** Resistance as a function of temperature at $v$=2 **(b)** and $D$=-0.4V/nm **(c)**. **(d)** $A_1$ and resistance as a function of displacement field at $v$=2. **(e)** numerical derivate $dR/dT$ as a function of displacement field and temperature at $v$=2.